\documentclass{aa}
\usepackage[varg]{txfonts}
\usepackage{graphicx}
\usepackage{natbib}

\newcommand{\g}{GS\,118+01$-$44}

\newcommand{\kms}{$\rm km\,s^{-1}$}
\newcommand{\am}{$^{\prime}$}

\newcommand{\hi}{\ion{H}{I}}
\newcommand{\hii}{\ion{H}{II}}

\begin{document}

\title{The \hi\, supershell \g\, and its role in the interstellar medium}
\author{L. A. Suad\inst{1},  S. Cichowolski\inst{2}, A. Noriega-Crespo\inst{3},  E. M. Arnal\inst{1,4}, J. C. Testori\inst{1}, \and N. Flagey\inst{5}}

\institute{Instituto Argentino de Radioastronom\'{\i}a (CCT-La Plata, CONICET), CC 5,
  1894, Villa Elisa, Argentina. 
  \and Instituto de Astronom\'{i}a y F\'{i}sica del Espacio (CONICET-UBA), Cuidad Universitaria, C.A.B.A, Argentina.
  \and Space Telescope Science Institute, 3700 San Martin Drive, Baltimore, DM 21218, USA.
  \and Facultad de Ciencias Astron\'omicas y Geof\'{\i}sicas, Universidad Nacional de La Plata, La Plata, Argentina.
  \and Canada-France-Hawaii Telescope Corporation, 65-1238 Mamalahoa Hwy, Kamuela, HI 96743, USA.
}

\date{received  /accepted  }

\abstract {}{We carry out a multiwavelength study to characterize the \hi\, supershell  designated \g, and to analyse its possible origin.}{A multiwavlength study has been carried out to study the supershell and its environs. We performed an analysis of the \hi, CO, radio continuum, and infrared emission distributions.}{The Canadian Galactic Plane Survey (CGPS) \hi\, data reveals that \g\, is centred at $(l, b) = (117\fdg7, 1\fdg4)$ with a  systemic velocity of $-44.3$ \kms. According to Galactic rotation models this structure is located at $3.0 \pm 0.6$ kpc from the Sun.  
There are several \hii\, regions and three supernova remnants (SNRs) catalogued in the region.
 On the other hand, the analysis of the temperature spectral index distribution shows that in the region there is a predominance  of  non-thermal emission. Infrared emission shows that cool temperatures dominate the area of the supershell.  Concerning the origin of the structure, we found that even though several OB stars belonging to  Cas\,OB5 are located in the interior  of \g, an analysis of the energy injected by these stars  through their stellar winds indicates that they do not have  sufficient energy to create \g. Therefore, an additional energy source is needed to explain the genesis of \g. On the other hand, the presence of several \hii\, regions and young stellar object candidates in the edges of \g\, shows that the region is still active in forming new stars.}{}


\titlerunning{Mutiwavelength study of  the \hi\, supershell \g.}
\authorrunning{L. A. Suad et al.}

\maketitle

\section{Introduction}\label{intro}

Although few
in number compared
with low-mass stars, massive stars, defined
as those stars with a main-sequence mass
of at least 8 M$_\odot$, are extremely important with regard to the physical and chemical conditions prevailing in the  interstellar medium  (ISM). Their importance stems from  their UV ionizing radiation and energetic winds, which  give rise to expanding \hii\, regions and interstellar bubbles.\   As a consequence of  stellar evolution, supernova explosions   take place within
these expanding features, and these supernovae further contribute to the expansion of   existing structures originating in previous stellar evolutionary stages. Since massive stars tend to be gregarious (e.g. as in an OB association), their cumulative effects may be one of the mechanisms at work, which  may give rise to large scale  structures known as supershells. Since these structures are mostly observed in the 21 cm line emission of the neutral hydrogen (\hi)  as regions of low \hi\, emission surrounded, totally or partially, by  walls of enhanced \hi\, emission, they are usually referred to as \hi\, supershells.

The first catalogue of Galactic \hi\, shells/supershells was published by \cite{hei79}. Later, based on different identification criteria, and making use of  diverse  \hi\, databases, several additional catalogues were elaborated  \citep{mcc02,ehl05,ehl13}. These \hi\, structures were also detected in nearby galaxies \citep{bag11}. Recently, using a combination of a traditional identification criteria and an automatic algorithm procedure, 566 \hi\, supershell candidates were identified in the second and third Galactic quadrants \citep{sua14}.

As the shells/supershells evolve in the ISM, the physical conditions for generating new stars can be fulfilled. Studies have been performed showing several cases of triggered star formation \citep[e.g.][]{arn07,sua12,cic14}.
Although \hi\, supershells are believed to be of importance in determining some large scale properties of the host galaxy (e.g. the gas injection into the lower Galactic halo), their origin and the role played by these structures in triggering star formation is a matter of debate.

In this paper, we present a comprenhesive study of one of the supershells detected in the catalogue of \cite{sua14}, designated as \g. We analyse the atomic, molecular, radio continuum, and  infrared emission in the region, trying to disentangle the origin of this
large structure and its interaction with the ISM .

\section{Observations}

High angular resolution \hi\, data were retrieved from  the Canadian Galactic Plane Survey  \cite[CGPS]{tay03}. Besides the high-resolution \hi\, data, the CGPS also provides high-resolution continuum data at 408 and 1420 MHz \citep{lan00}.
The CGPS data base also comprises other data sets that have been reprojected and regridded to match the DRAO images. Among them is the Five College Radio Astronomical Observatory (FCRAO) CO Survey of the Outer Galaxy \citep{hey98}.

Mid-infrared emission at 12 and 22 $\mu$m were obtained from the Wide-field Infrared Survey Explorer \citep[WISE;][]{wri10}. These data were taken from the Nasa/IPAC Infrared Science Archive (IRSA) home page\footnote{http://irsa.ipac.caltech.edu/}. 

Far-infrared emission at 65, 90, 140, and 160 $\mu$m were retrieved from the AKARI satellite \citep{tak15}.
We  also used Herschel (PACS) 160 $\mu$m data (red channel) \citep{mol10} and Planck 350 and 550 $\mu$m data \citep{pla15}.
In Table \ref{tabla-observations} all the observational parameters are listed.

\begin{table}
\caption{Observational parameters.} 
\label{tabla-observations}
\centering
    \begin{tabular}{l c}
 \hline\hline
\textbf{CGPS \hi\, data} &   \\
Angular resolution & $1\farcm2 \times 1\arcmin $ \\
Velocity resolution & 1.3 \kms \\
Channel separation & 0.82 \kms \\
Velocity coverage & --164.7 to 58.7 \kms \\
Observed rms noise & 1.5 K \\
\hline
\textbf{Radio continuum} & \\
Angular resolution (408 MHz) & 3.2\arcmin x 2.7\arcmin \\
Observed rms noise  & 0.4 K\\
Angular resolution (1420 MHz) & 1\arcmin x 1.2\arcmin \\
Observed rms noise & 0.03  K\\
\hline
\textbf{Infrared data} & \\
 \textbf{WISE} (3.4, 4.6, 12, 22 $\mu$m)& \\
Angular resolution& 6\farcs1 -- 12\arcsec\\
\textbf{AKARI} (65, 90, 140, 160  $\mu$m) & \\
Angular resolution & 63\arcsec -- 88\arcsec \\
\textbf{PACS-}160$\mu$m & \\
Angular resolution & 13\arcsec \\

\textbf{Planck} (350, 550 $\mu$m) & \\
Angular resolution & 5\arcmin \\
\hline
\end{tabular}
\end{table}

\section{HI emission distribution}

\subsection{Previous H{\sc i} works towards $l$  $\sim$ 118$^\circ$}\label{previous}

\cite{fic86}, using the \hi\, database of  \cite{wea73}, reported an \hi\, supershell likely to be associated with three Galactic supernova remnants.
This feature, centred at $(l, b)$ = (117\fdg5, 1\fdg5), is detected 
along the velocity range from $-60$ to $-35$ \kms, and is ellipsoidal in shape with a major and minor axis of  about 7$^\circ$ and $\sim$ 3$^\circ$ in Galactic longitude and latitude, respectively.

Later, \cite{moo03}, using the Leiden-Dwingeloo \hi\, survey \citep{har97}, detected a large \hi\, circular  feature with an angular diameter of  $\sim$8\fdg5, covering the velocity range from $-60$ to +3 \kms\,  and centred at 
$(l, b)$ = (117\fdg5, 1\fdg0). The authors suggest that this structure was created by the interaction of the massive stars of Cas\,OB5 with their surrounding ISM.

Later on, \cite{cic09}, using the CGPS  database \citep{tay03},  reported that the \hii\, region Sh2-173 was  born and is evolving onto the borders  of a large elliptical feature (whose major and minor axis have angular diameter of 6\fdg1 and 3\fdg9, respectively), which is centred at $(l, b)$ = (117\fdg8, 1\fdg5). This large feature,  labelled by \cite{cic09} as GSH117.8+1.5-35, covers the velocity range from $-50$ to $-20$ \kms, and according to the authors it may have also been created by the action of  early-type stars of the OB-association Cas\,OB5 on the surrounding medium.

Finally,  an elliptical feature, labelled as \g\, in the catalogue of \cite{sua14}, with its geometrical centroid at $(l, b)$ = (117\fdg9, 1\fdg2) and a barycentric velocity of $-44.3$ \kms\, is listed. The angular diameter of its major and minor axis are 
5\fdg0 and 3\fdg8, respectively, and the structure is visible along the velocity range from $-52$ to $-35$ \kms.

\begin{figure}
\includegraphics[scale=0.8]{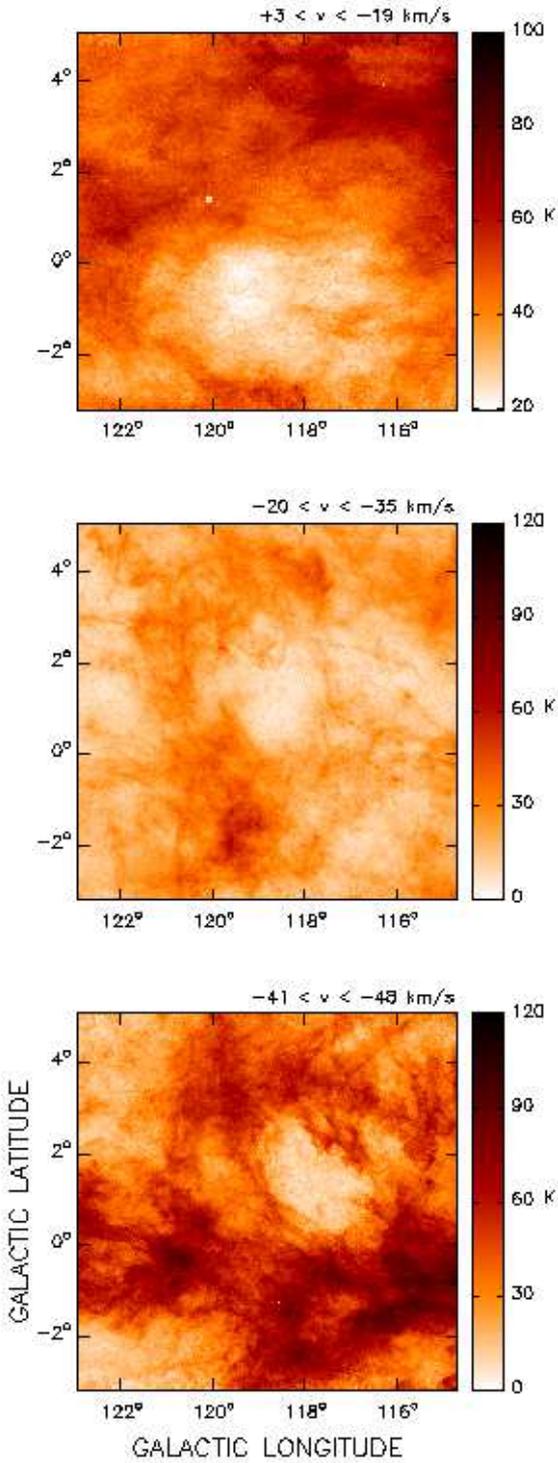}
\caption{Mean brightness temperature of the CGPS \hi\, emission distribution between the velocity range from: \textit{upper panel:} +3 to $-19$ \kms, \textit{middle panel:} $-20$ to $-35$ \kms, and \textit{lower panel:} $-41$ to $-48$ \kms.}
\label{3gs}
\end{figure}

Summing up, it is clear that the region around Cas\,OB5 presents  several \hi\, structures that deserve further careful analysis. Figure \ref{3gs} shows three images that reveal the  spatial CGPS \hi\, emission 
distribution along most of the velocity range covered by the \hi\, feature identified by \cite{moo03}.
In all three images a minimum in the \hi\, distribution, surrounded by walls of enhanced \hi\, emission is clearly 
visible, revealing a shell-like structure. It is worth mentioning,  however, that at velocities above $-19$ \kms\, (\textup{\textit{\textup{top panel}}} in Fig. \ref{3gs}) the shell's barycentre is detected at ($l,b$) $\sim (119^\circ, -1^\circ)$,
while at  velocities lower than $-20$ \kms\, the shell's centre is located  above the Galactic plane (see Fig. \ref{3gs}, \textit{\textup{middle and lower panels}}). This noticeable and sudden angular shift in the location of the shell is hard to reconcile with the idea that we are observing a single \hi\, structure through the entire velocity range shown in the three images  as was stated by \cite{moo03}.
Therefore, our first conclusion is that the \hi\, feature observed along the velocity interval from $-19$ to +3 \kms\, is different, and does not necessarily have a physical  connection with the \hi\, feature observed between $-20$ and $-48$ \kms.

\begin{figure*}
\includegraphics[scale=0.8]{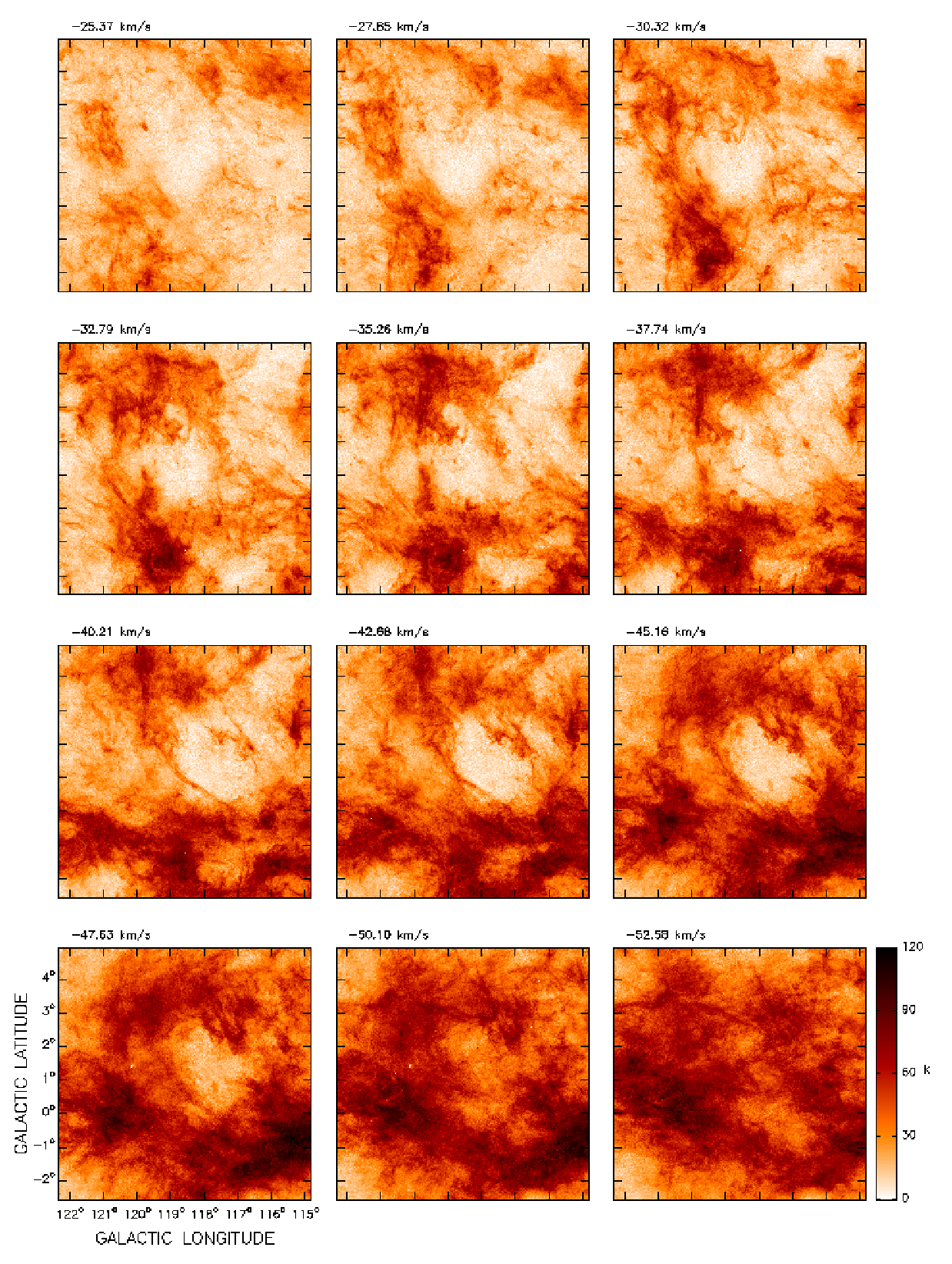}
\caption{CGPS neutral hydrogen emission in the environs of  \g. Each panel shows the mean brightness temperature within three consecutive velocity channels. At the top of each panel the central velocity is indicated.}
\label{12channels}
\end{figure*}

\begin{figure}
\includegraphics[angle=-90,scale=0.34]{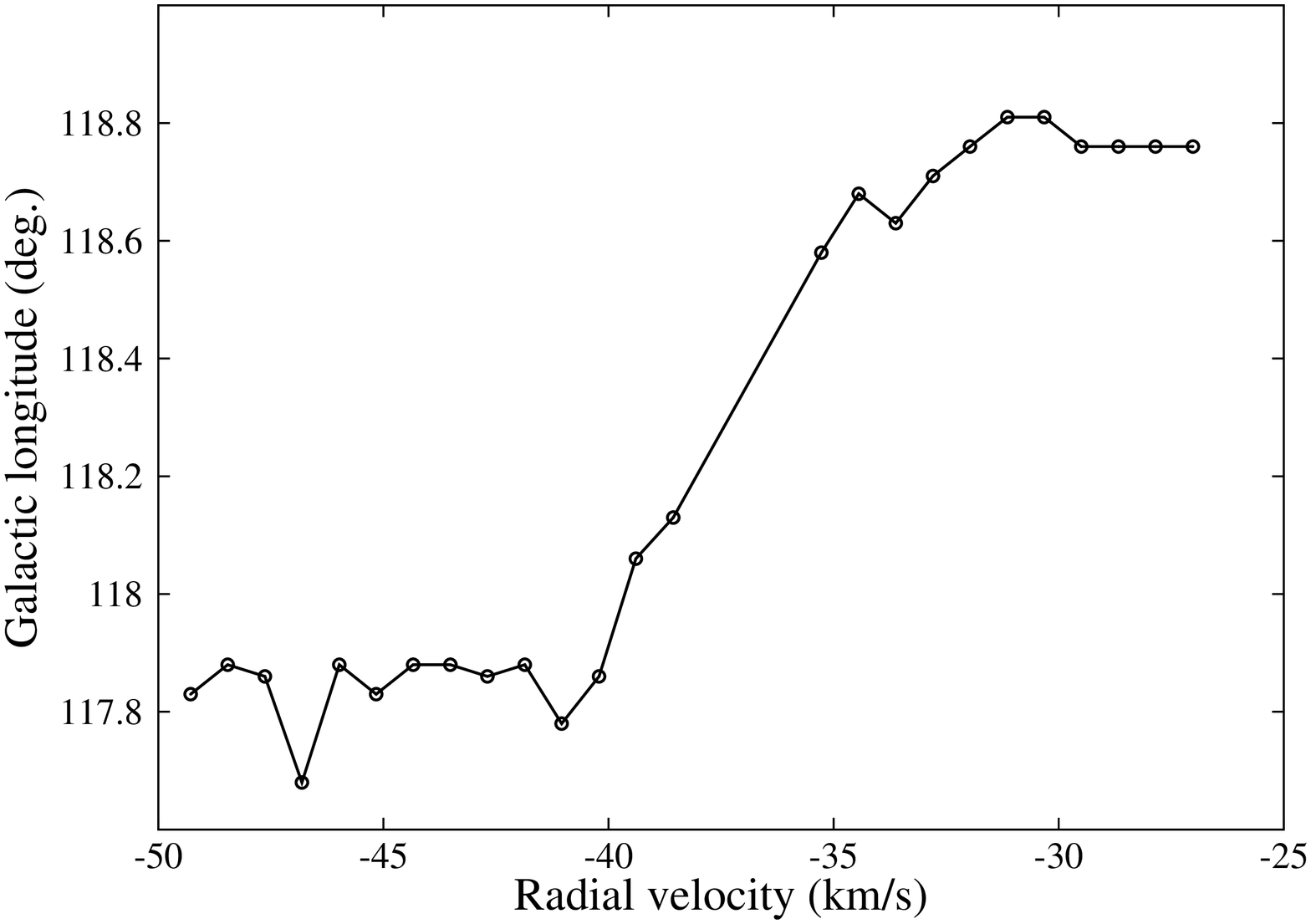}

\vspace{0.7cm}

\includegraphics[angle=-90,scale=0.34]{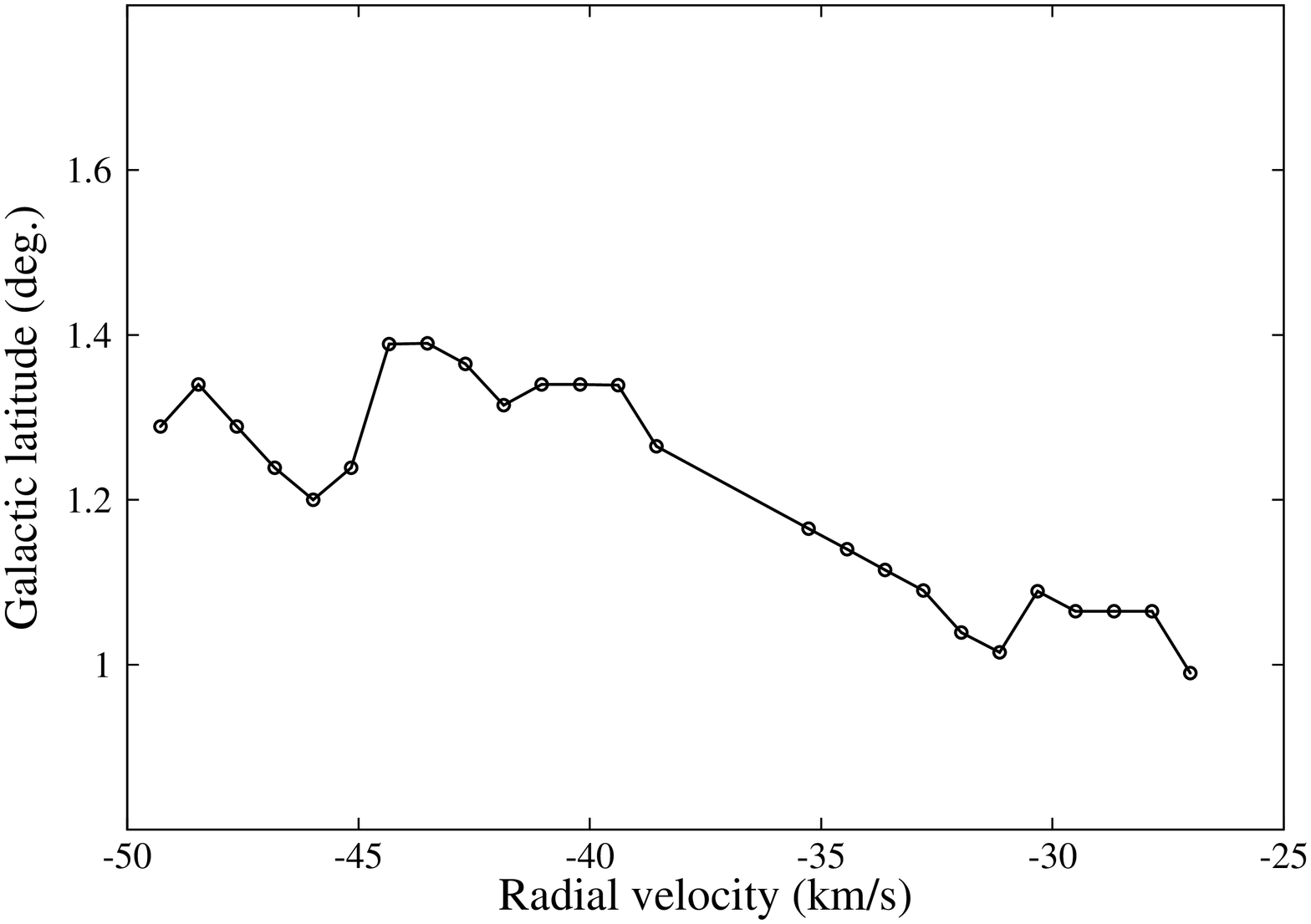}
\caption{Baricentral coordinates of the two structures observed at different radial velocities. \textit{Upper panel:} Galactic longitude baricentral coordinates versus  radial velocity. \textit{Lower panel:} Galactic latitude baricentral coordinate versus radial velocity.} 
\label{2centers}
\end{figure}

To better analyse the area regarding velocities below -20 \kms, Fig. \ref{12channels}  shows 12 images covering the velocity range from $-53.4$ to $-24.55$ \kms. Each image shows the mean brightness temperature of three consecutive CGPS velocity channels. Though the \hi\, minimum shown in Fig. \ref{3gs} (\textit{\textup{middle} \textup{and lower panels}}) is clearly observable in most of the images, a closer look at them shows that
the barycentral coordinates of the \hi\, minimum detected at more positive
velocities are slightly shifted from those observed at more negative
velocities. This is better seen in Fig. \ref{2centers}, where the Galactic coordinates of the \hi\, shell's barycentre,  as a function
 of the radial velocity, are shown.  The barycentric coordinates have been calculated by fitting an ellipse for different radial velocities, whose input points are the maxima of \hi\, emission surrounding the \hi\, voids. The uncertainty in the centroid is typically $\pm 0\fdg1$.  Strikingly enough, the barycentric Galactic longitude of the \hi\, minimum differs  by $\sim$1$^\circ$ between $-35$ $\leq$ {\it v} $\leq$
 $-25$ \kms\, and $-50$ $\leq$ {\it v} $\leq$ $-40$ \kms. A shift of about $0.3^\circ$ is also observed in the feature's centroid at Galactic
 latitude. 
 Thus, we believe that along the velocity interval  $-50 \leq v \leq -25$ \kms\, we are dealing with {\it two different} Galactic features and not just with one single structure.

The feature observed at more negative velocities is  catalogued as \g\, by \cite{sua14}.  This feature has a good morphological correspondence with that  found by \cite{fic86}. From here
 onwards we  concentrate our study on this feature.

\subsection{\g}\label{g}

\begin{figure}
\includegraphics[angle=-90,scale=0.34]{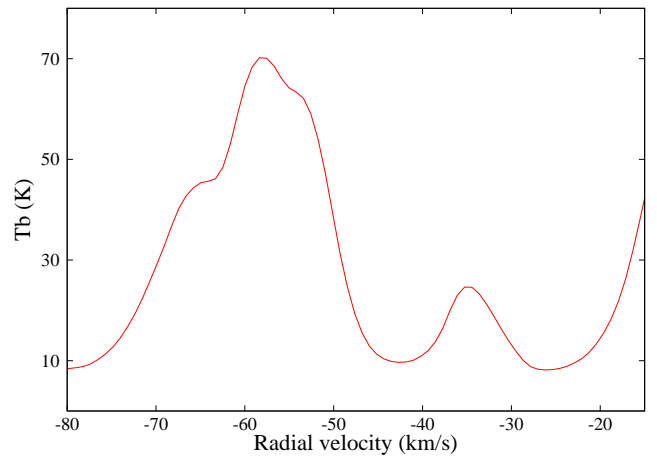}
\caption{Averaged \hi\, profile in a $0\fdg5 \times 0\fdg5$ region located at the centre of \g.}
\label{average-profile}
\end{figure}

An inspection of Fig. \ref{12channels} clearly shows the presence of a well-defined local \hi\, minimum in brightness temperature that we identify as \g. This feature is  centred at $(l, b) \sim (117\fdg9, 1\fdg2)$. To better determine the velocity interval where this structure is seen, an averaged profile has been derived within a $0\fdg5 \times  0\fdg5$ box centred at $(l, b) = (117\fdg9, 1\fdg2)$  along the velocity range from $-15$ to $-80$ \kms\, (see Fig. \ref{average-profile}). There \g\, is detected between the peak of emission located at about $-35.0$ \kms\, and the shoulder detected at about -52.3 \kms, which corresponds to the \hi\, emission seen at -52.6 \kms\, in the central position of \g\, in Fig. \ref{12channels}. On the other hand, the emission observed at $-35.0$  \kms\, is associated with the emission detected at $(l, b) \sim (117\fdg9, 1\fdg2)$ in the \textit{\textup{middle panel}} of Fig. \ref{3gs} . From Fig. \ref{average-profile} it can be inferred that the systemic velocity  ($v_0$, where the feature attains its maximum angular extent) of \g\, is $-44 \pm 2$ \kms.

\begin{figure}
\includegraphics[scale=0.35]{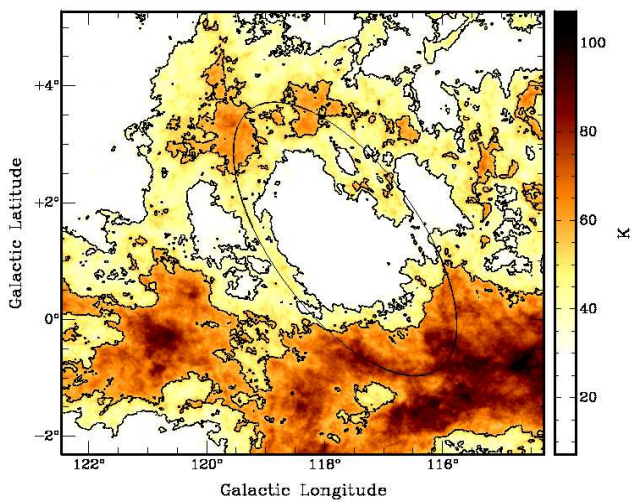}
\caption{Mean brightness temperature of the \hi\, emission distribution associated with \g\, in the velocity range from -41.9 to -47.6 \kms. Contour levels are at 35  and 55 K. Angular resolution is  2\am.}
\label{HIproV-auto}
\end{figure}

The mean brightness temperature distribution in the velocity range from $-41.9$ to $-47.6$ \kms\, is shown in Fig. \ref{HIproV-auto}.  Since the main physical parameters given for \g\, in the catalogue of \cite{sua14} were obtained using the Leiden-Argentine-Bonn (LAB) survey \citep{kal05} (FWHM 30\arcmin), we now use the CGPS data available for this structure to recalculate them.  To this end, an ellipse has been fitted to the \hi\, shell with the classical minimum mean square error (MMSE) technique \citep{hal98}. Every point used in this fitting corresponds to the local  maximum brightness temperature that surrounds the \hi\, minimum. The obtained ellipse is shown superimposed in Fig. \ref{HIproV-auto}, where we indicate its  centroid $(l_e, b_e)= (117\fdg7 \pm 0\fdg1, 1\fdg4 \pm 0\fdg1)$, its  major and minor semi-axes $a= 2\fdg7 \pm 0\fdg3$ and $b = 1\fdg2 \pm 0\fdg1$, respectively, and the inclination angle ($\phi$) of the semi-major axis with respect to the Galactic plane, $\phi =  126^\circ \pm 4^\circ$ (measured counterclockwise from the Galactic plane).

According to the systemic velocity  of the structure, and taking the non-circular motions present in this region into account \citep{bra93}, we estimated a distance to the Sun of $3.0 \pm 0.6$ kpc for \g. At this distance, the major and minor semi-axes are  $140 \pm  30$ and $60 \pm 10$ pc, respectively, and the effective radius of the structure   is $R_{ef} = \sqrt{a\,b} = 94 \pm 15$ pc.

To determine the total gaseous mass associated with the supershell, we  used $M_{\rm {HI}}=N_{\rm {HI}}A_{\rm {HI}}$, 
where  $N_{\rm {HI}}$ is the \hi\, column density, $N_{\rm {HI}}=C\int^{v_1}_{v_2} T_{b} \, dv $, $v_1=-35.0$ \kms\,  and $v_2=-52.3$ \kms\, are the velocity interval, where the structure is detected and $ T_{b}$ the brightness temperature.
The parameter $A_{\rm {HI}}$ is the area of the supershell, $A_{\rm {HI}}=\Omega_{\rm {HI}}d^2$, where $\Omega_{\rm {HI}}$ is the solid angle covered by the structure and $d$ is the distance to the Sun.
Adopting solar abundances, we estimated a total gaseous mass of  $M_{\rm {t}} = 1.34\,M_{\rm {HI}}= (4.9 \pm 2.2)\times 10^5 \,M_{\odot}.$

The dynamic age of the structure is defined as $t_{dyn} = R_{ef}/v_{exp}$, where $v_{exp}$ is the expansion velocity of the feature ($v_{exp} = \Delta v/2$, where $\Delta v$ is the velocity range where the structure is observed).
Adopting  $v_{exp} = 8.7 \pm 1.6$ \kms, we obtained $t_{dyn}  = 10.8 \pm 2.6$ Myr.

\section{CO emission }

\begin{figure}
\centering
\includegraphics[width=9cm]{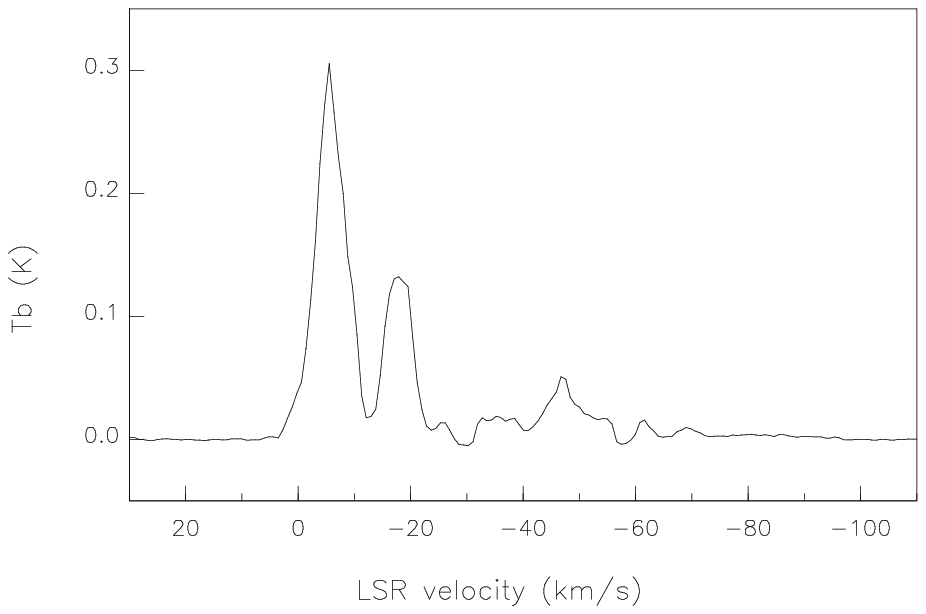}
\includegraphics[width=9cm]{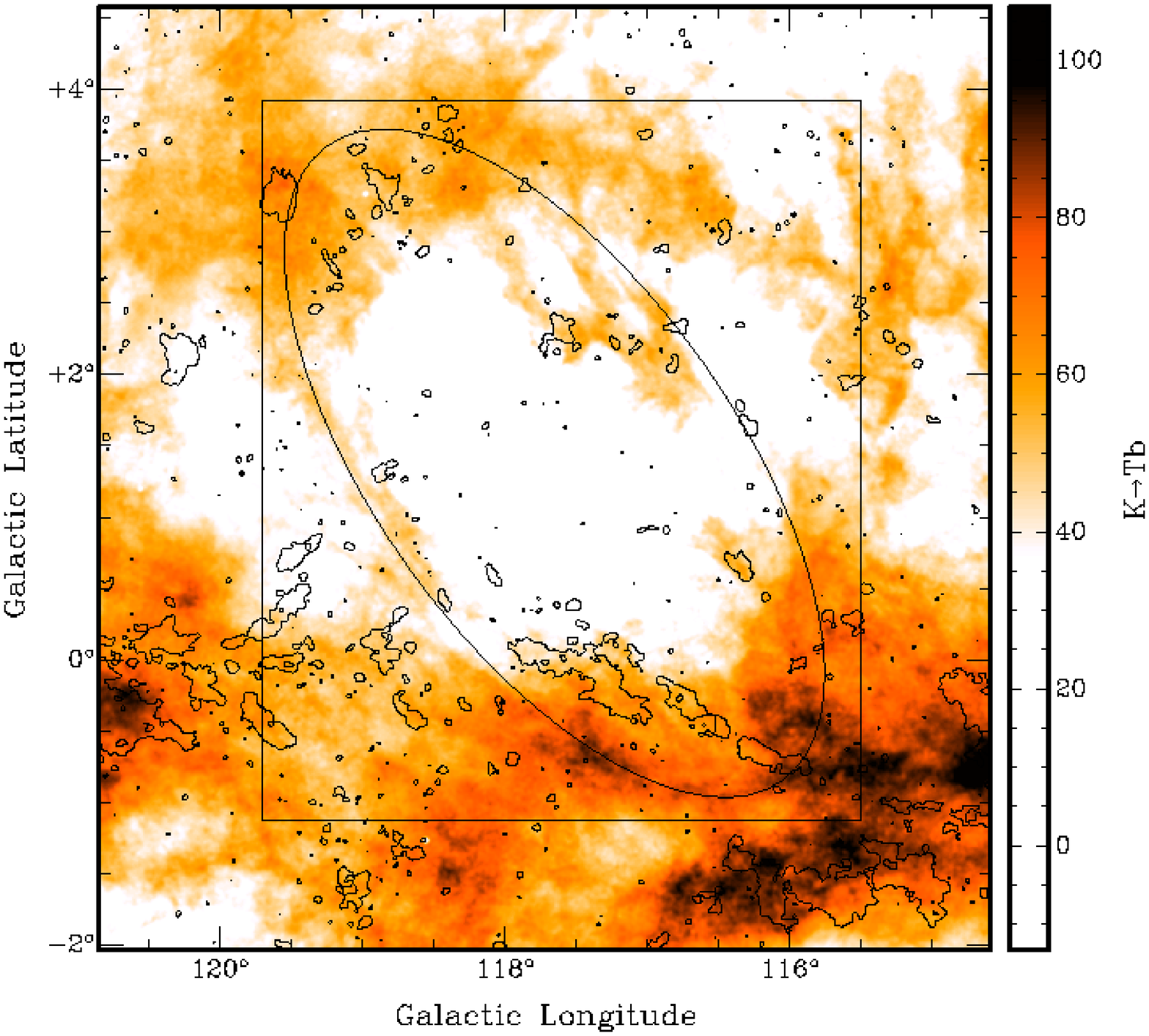}
\caption{\textit{Top panel}: mean CO emission within a rectangular area 4\fdg2 $\times$ 5\fdg05 in size, centred at ($l, b$) = ( 117\fdg6, 1\fdg4). \textit{Bottom panel}: mean brightenss temperature of the CGPS \hi\, emission distribution in the velocity range from $-41.9$ to $-47.6$ \kms. The contour corresponds to the 1.7 K level of the CO  emission  integrated in the velocity interval from --53 to --40 \kms. The ellipse is that fitted for the \hi\, data and the box indicates the area used to obtain the CO profile shown in the \textit{top panel}.}
\label{co}
\end{figure}

In this section, we analyse the CO emission distribution  looking for molecular clouds possibly related to \g.
 A CO profile depicting the mean molecular emission originated within a rectangular region of 4\fdg2 $\times$ 5\fdg05 in size and centred at ($l, b$) = ( 117\fdg6, 1\fdg4) is shown in the top panel of Fig. \ref{co}. As can be seen from this profile, although in this region the bulk of the molecular emission arises from the velocity interval between 5 and --25 \kms, an additional, smaller peak is detected in the velocity range from about $-40$ to $-53$ \kms. Taking into account the systemic velocity estimated for the \hi\, structure, $v_0 = -44$ \kms, we infer that this emission may stand a chance of being associated with \g. 
In the bottom panel of Fig. \ref{co}, the \hi\, emission distribution averaged as in Fig. \ref{HIproV-auto} is shown with the CO emission present in the velocity range from $-53$ to $-40$ \kms\,  indicated by contours.
The presence of several small and disperse CO clouds around \g\, can be observed.

 It is difficult to determine whether there is a relationship among these small molecular clouds and the \hi\, supershell. Moreover, the origin of these molecular features is not clear. Either they are part of a pre-existing molecular cloud, which was dissociated and fragmented by the action of the massive stars or they were created in the inner parts of the dense  \hi\, wall, as proposed by \citet{daw11, daw15}.

To characterize these molecular features, we  calculated their molecular masses using the relationship $ N(\rm{H_2}) = X\, W_{\rm{CO}}$, where $N(\rm{H_2})$   is the H$ _2$ column density, $X$ is the CO-to-H$_2$ conversion factor, $X = 1.9 \times 10^{20}$ cm$^{-2}$ (K \kms)$^{-1}$ \citep{gre90,dig95}, and $W_{\rm{CO}} = \int T_b(\rm{CO})\,dv$ is the integrated CO line intensity over the velocity range from $-53$ to $-40$ \kms.
Then, the molecular mass of each cloud was derived from $M_{\rm{H}_2}[{\rm{M}}_\odot] = 4.2 \times 10^{-20}\, N({\rm{H}}_2)\, d^2\, A$, where $d$ is the distance in pc and $A$ is the area in steradians.
 The estimated mass of the small clouds ranges from around 15 to 1.6 $\times 10^4$ M$_\odot$, and 
summing all the observed structures, we obtained a total molecular mass of $M_{\rm{H_2}}=(7.4 \pm 3.2)\times 10^4$ M$_\odot$,  where the uncertainty arises mainly from the area estimates and the adopted distance. This amount of mass is just the 20\% of the \hi\, mass inferred to be present in the shell.

\section{Radio continuum}

\begin{figure}
\includegraphics[scale=0.35]{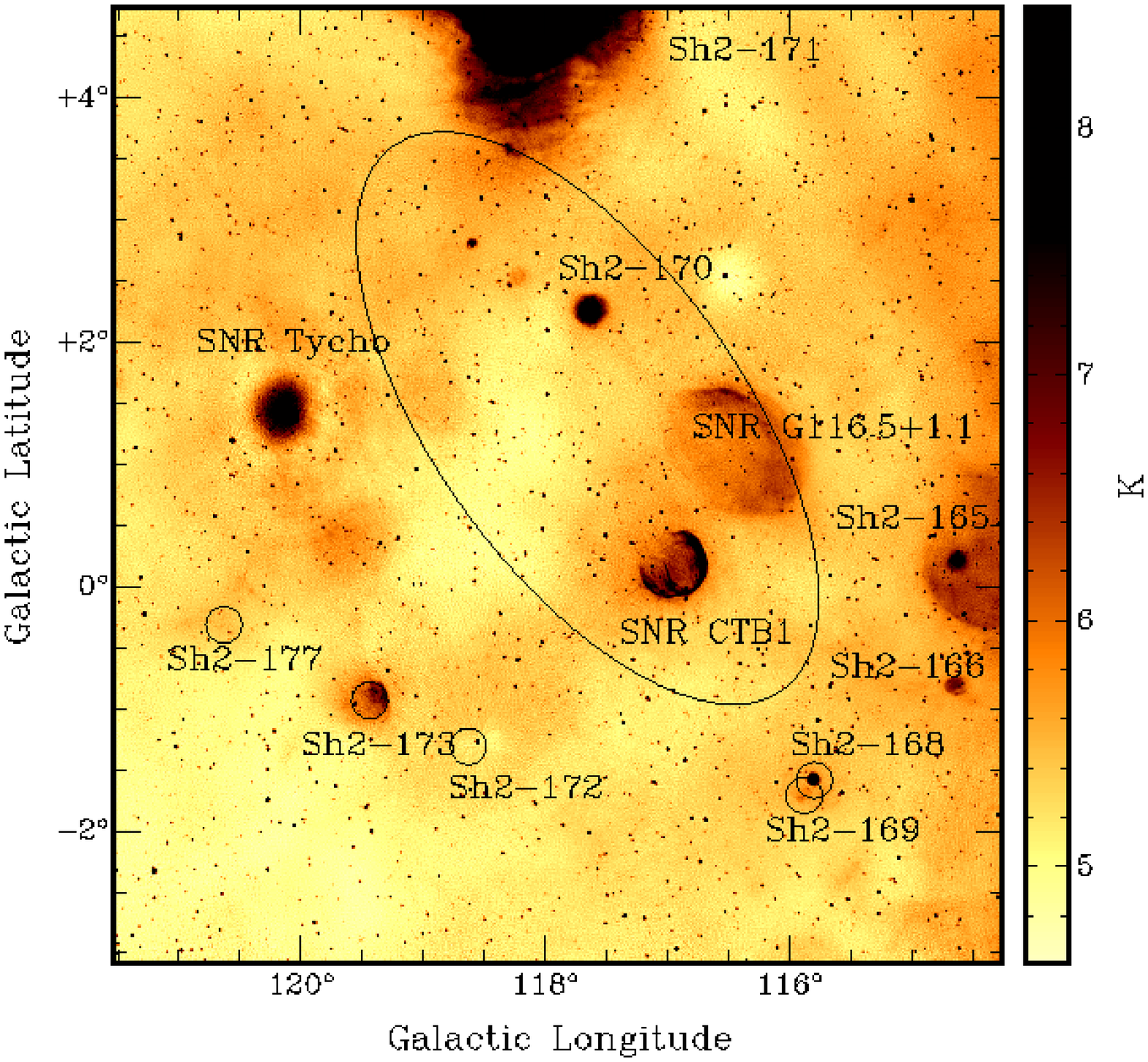}
\includegraphics[scale=0.35]{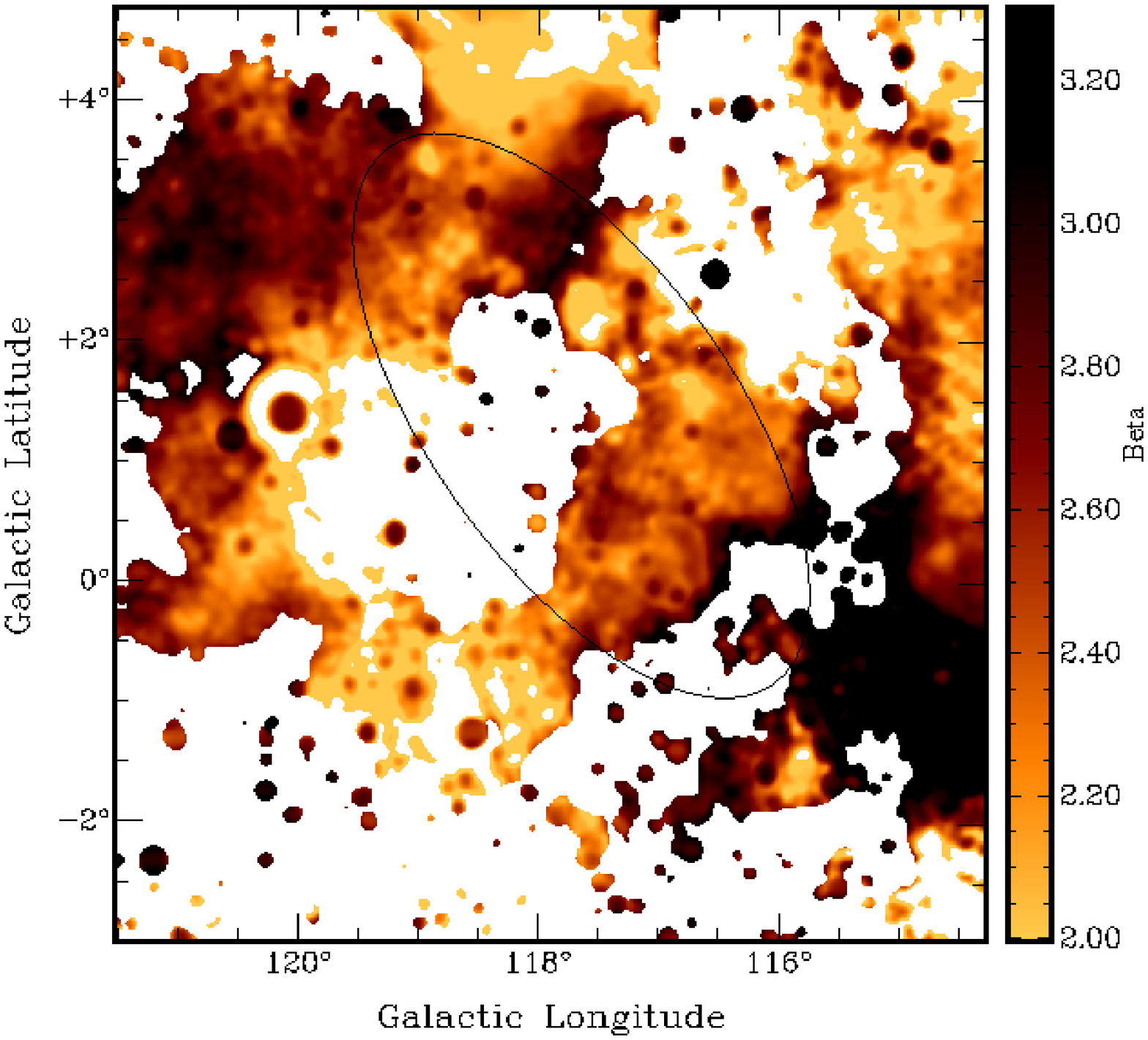}
\caption{\textit{Upper panel}: CGPS radio continuum emission at 1420 MHz in the area of \g. The Sharpless \hii\, regions and SNRs present in the region are labelled. \textit{Lower panel}: temperature spectral index map. The ellipse delineates the \hi\, supershell location.}
\label{1420-RHII}
\end{figure}

We now study the distribution of the radio continuum emission in the region of \g\, using CGPS data at 408 and 1420 MHz.   Figure \ref{1420-RHII} (\textit{\textup{upper} \textup{panel}}) shows the emission at 1420 MHz, with the ellipse delineating the location of \g\, superimposed.  The presence  of several bright and extended regions, already  catalogued as \hii\, regions  and  SNRs,  are evident. According to the WISE Catalog of Galactic HII regions\footnote{http://astro.phys.wvu.edu/wise/} \citep{and14}, the \hii\, regions Sh2-165, Sh2-168, Sh2-169, and Sh2-172 have radial velocities in the range from $-42.6$ to $-47.9$ \kms, which coincides with the radial velocity interval where \g\, is detected. Given the relative location among the mentioned H\,II regions and \g\, in the plane of the sky and in velocity, we cannot discard the possibility that these structures were related to each other. On the other hand, three SNRs are observed in this area, namely G116.5+1.1, CTB1, and Tycho. The estimated distances of the SNRs shown in Fig. \ref{1420-RHII} ({upper panel}) are 1.6 kpc for G116.5+1.1 and CTB 1 \citep{uya04}, and  2.4 kpc for Tycho  \citep{lee04}. Although Tycho is the  nearest to \g, it is located outside the borders of the supershell. Therefore, these  SNRs are very likely not associated with \g. 

Besides the HII regions and SNRs,  Fig. \ref{1420-RHII} ({upper panel}) shows a minimum in
the radio continuum emission coincident with the interior part of the \hi\, shell, while diffuse emission is detected around it.
To carry out a  study of the nature of the diffuse continuum emission, we studied the general distribution
of the temperature spectral index. The spectrum of radio continuum radiation is usually described by the temperature 
spectral index $\beta$ ($T_b \sim  \nu ^{-\beta}$, where $T_b$ is the brightness temperature at the frequency $\nu$). The flux spectral index $\alpha $ is related to $\beta$ as $\alpha = \beta-2$. 

To analyse the distribution of the temperature spectral index across the map, an adjustment
of the adopted Galactic zero levels of the different survey frequencies, and the overall Galactic 
background continuum emission have to be subtracted out. To this aim we applied 
the background filtering method (BGF) developed by \cite{Sof79}. A 4$^\circ$x4$^\circ$ degree filtering beam was applied to each 
frequency. The DRAO data at 408 MHz and 1420 MHz were first smoothed by a Gaussian beam to the angular resolution of 6\arcmin x 6\arcmin. The corresponding  final rms of these images are $\sigma_{\rm{1420}} = 0.4$ K at 1420 MHz and $\sigma_{\rm{408}} = 0.03$ K at 408 MHz. 

The temperature spectral index between two frequencies $\nu_1$ (408 MHz) and $\nu_2$ (1420 MHz) is calculated pixel wise from
$$\beta = log(T_b(\nu_1)/T_b(\nu_2))/log(\nu_1/\nu_2),$$ 
\noindent where we used values for $T_b$ greater than 3$\sigma$ for both frequencies. The resulting temperature spectral index map  is shown in Fig. \ref{1420-RHII} ({lower panel}). This map shows a spectrum that is a mixture of 
synchrotron emission and thermal emission. 
Flat spectrum areas ($\beta=2.1$), where thermal emission is a significant component, were only found towards the eastern region of \g\  between $118\fdg7 \leq l \leq 120\fdg3$ and $-0\fdg1 \leq b \leq 1\fdg9$. 
The values for the non-thermal emission mostly run between $2.3 \leq \beta \leq 2.7$ and dominate the remaining parts of the shell.
The errors, stemming from the uncertainty in determining the background levels, are of the order of  $\Delta \, \beta = 0.1$.

\section{Infrared emission}

\begin{figure}
\includegraphics[scale=0.35]{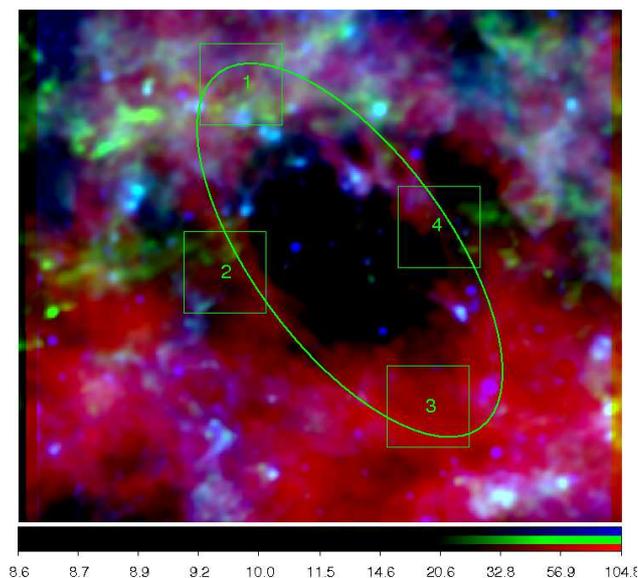}
\caption{Colour-composite image of the area of \g. Blue 
shows the emission at 12  $\mu$m (W3 WISE band), green  represents the
emission at 550  $\mu$m (Planck), and red shows the mean brightness temperature of the CGPS \hi\, emission distribution between the velocites -41.9 and -47.6 \kms. The ellipse superimposed is that fitted for the CGPS \hi\, data and the square regions delimit the zones where the fluxes have been calculated.}
\label{12-550um-HI}
\end{figure}

\begin{figure}
\includegraphics[scale=0.35]{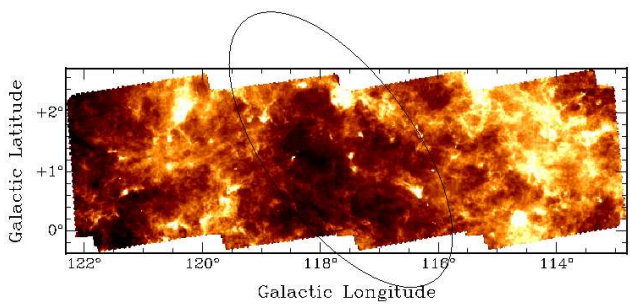}
\caption{Infrared emission distribution at 160 $\mu$m in a partial region of \g. The ellipse superimposed is that fitted for the CGPS \hi\, data. Regions with less infrared emission are shown as dark regions.}
\label{herschel1-red}
\end{figure}

At mid- and far-infrared wavelengths the region occupied by the HI
   supershell shows a complex morphology, although there is clearly a
   minimum of IR emission at the centre and some compact
   enhancement at its borders, i.e. it mimics some of the structure
   defined by the HI shell.  This spatial distribution of the IR
   emission is detected at wavelengths as short as  12 $\mu$m (WISE)
   and as long as 550 $\mu$m (Planck). Figure \ref{12-550um-HI}  shows a three-colour image with WISE  12 $\mu$m in blue, Planck 550 $\mu$m in green, and the CGPS \hi\, emission  averaged between $-41.9$ and $-47.6$ \kms\, in red, with an overlay of the ellipse
   fitted in the HI data centred at $(l, b) \sim (117\fdg8, 1\fdg2)$.
   The available Herschel data from the HiGAL Survey
   \citep{mol10} does not entirely cover  the \g\,
   region, as shown in the 160 $\mu$m map (see Fig. \ref{herschel1-red}).

\subsection{The dust properties}

One of the questions that  can be addressed with the mid- and far-infrared continuum data is
that of the properties of the dust in the shell, such as its temperature and mass.
For this analysis, we  followed the standard technique using dust models to match
the mid- to far-IR spectral energy distribution (SED); and based on the best fits to constrain
the dust properties. We used the recently release DUSTEM \citep{com11} 
models that use a combination of three dust components, polycyclic aromatic hydrocarbons (PAH), very small (VSG) and big dust grains (BGs), to interpret the emission from dust between $\sim 3$ to 1000 $\mu$m\,  \citep[see e.g.][]{li01,com10,com11}. 
Currently the available version of DUSTEM is missing the necessary
  database to include the WISE bands in the modelling. This is in part because the Atlas WISE images are not delivered in flux density units
  (Compiegne, private communication). 

\begin{table*}
\caption{Flux densities estimated for regions 1, 2 , 3, and 4.} 
\label{fluxes}
\centering
\begin{tabular}{c c c c c c c c c}
\hline\hline
& & & & Fluxes (Jy) & & & &\\
Region & ($l$, $b$) & 65 $\mu$m & 90 $\mu$m &  140 $\mu$m & 160 $\mu$m & 350 $\mu$m & 550 $\mu$m\\
\hline
1  & $(119\fdg0, 3\fdg5)$& 2890 & 7192 &  22013 & 22350 & 12739 & 4474 \\
2 & $(119\fdg2, 1\fdg2)$ & 1584  &   3283   &  11358   & 10838   &  5253   &  2020\\
3 & $(116\fdg7, -0\fdg5)$ & 1340  &    2263 &  6715  &  4973  & 1073 &  358 \\
4 & $(116\fdg6, 1\fdg7)$ & 1733 &  3339  &  9553  & 8161  & 3967  &  1480\\

\hline
\end{tabular}
\end{table*}

 To characterize the dust properties, we  selected four regions ($1^\circ$ x $1^\circ$) located at the edge of the shell, as indicated in Fig. \ref{12-550um-HI}. The integrated flux densities for these regions were estimated
by, first, convolving with a 5\arcmin\, Gaussian beam,  to match the
lower resolution of the 550 $\mu$m Planck image, all the images, i.e. 65, 90, 140, and 160 $\mu$m (from AKARI), plus 350 $\mu$m (from Planck). And, second, via an On and Off  over
the shell measurements; this last image was taken at the centre of the supershell and was subtracted from the On value to try to mitigate the effect of the foreground emission contamination, which  is expected to be present on most line-of-sight measurements. The estimated flux densities are listed in Table \ref{fluxes}.

\begin{figure*}
\includegraphics[scale=0.35]{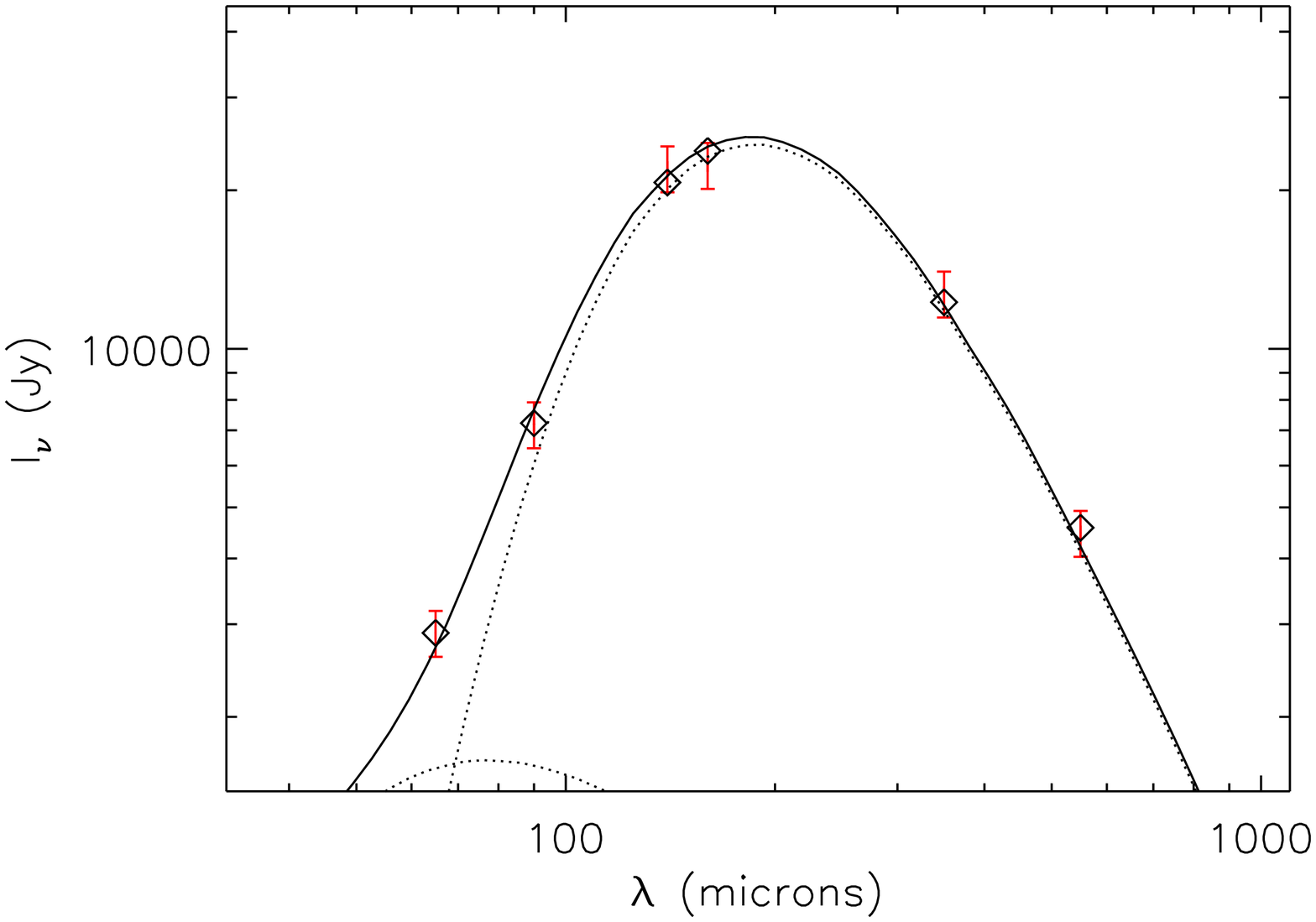}\includegraphics[scale=0.35]{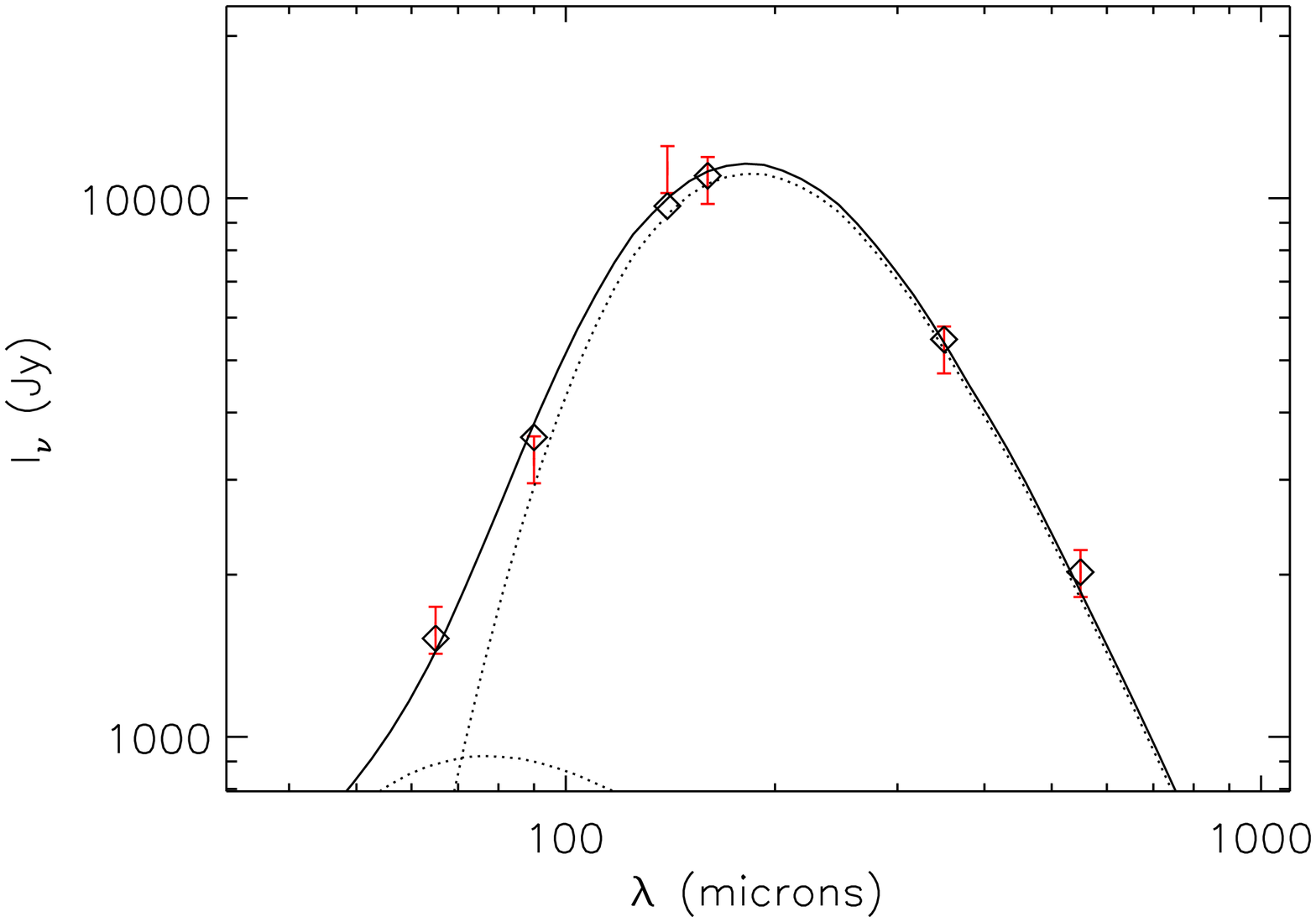}
\vspace*{0.5cm}
\includegraphics[scale=0.35]{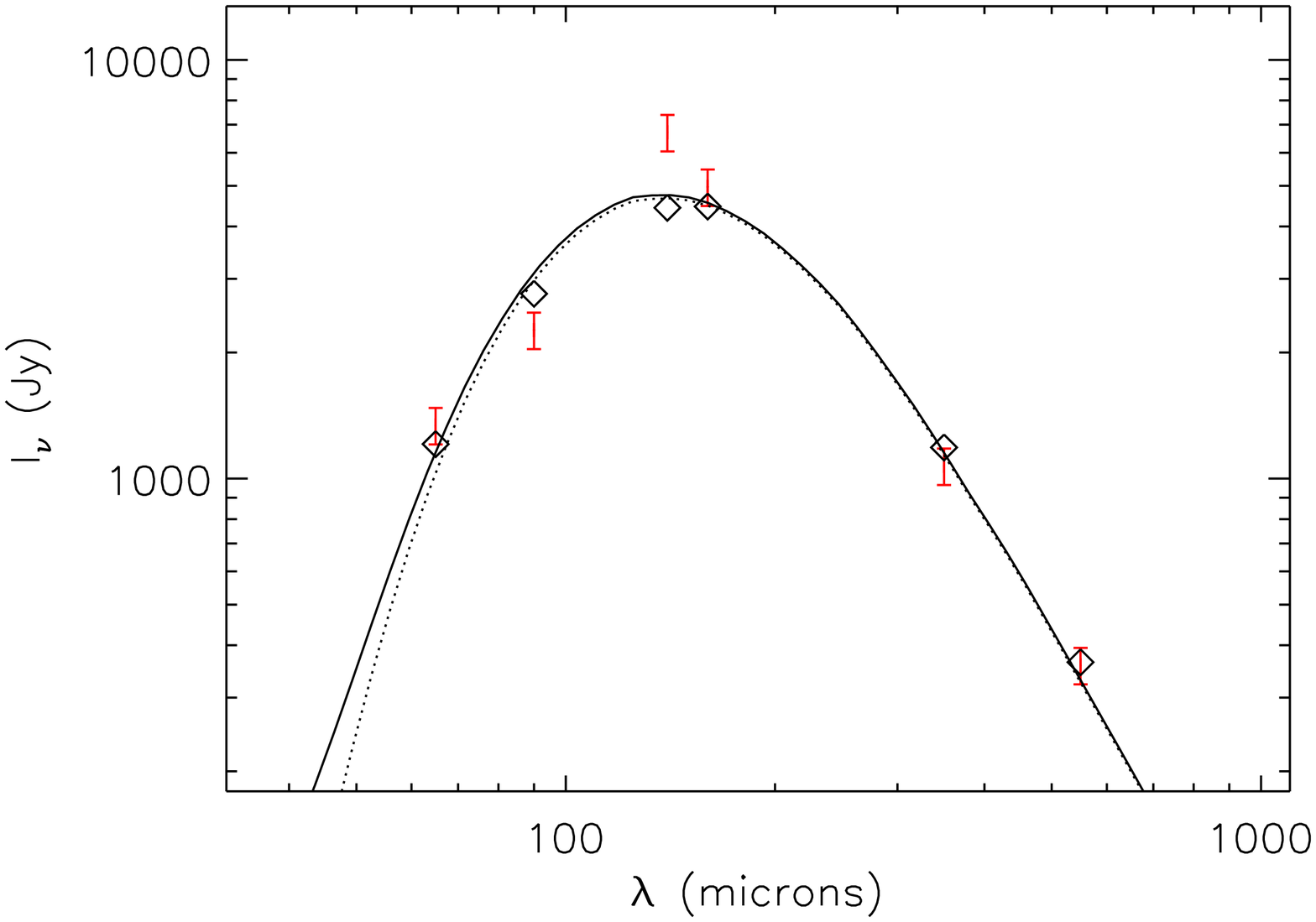}\includegraphics[scale=0.35]{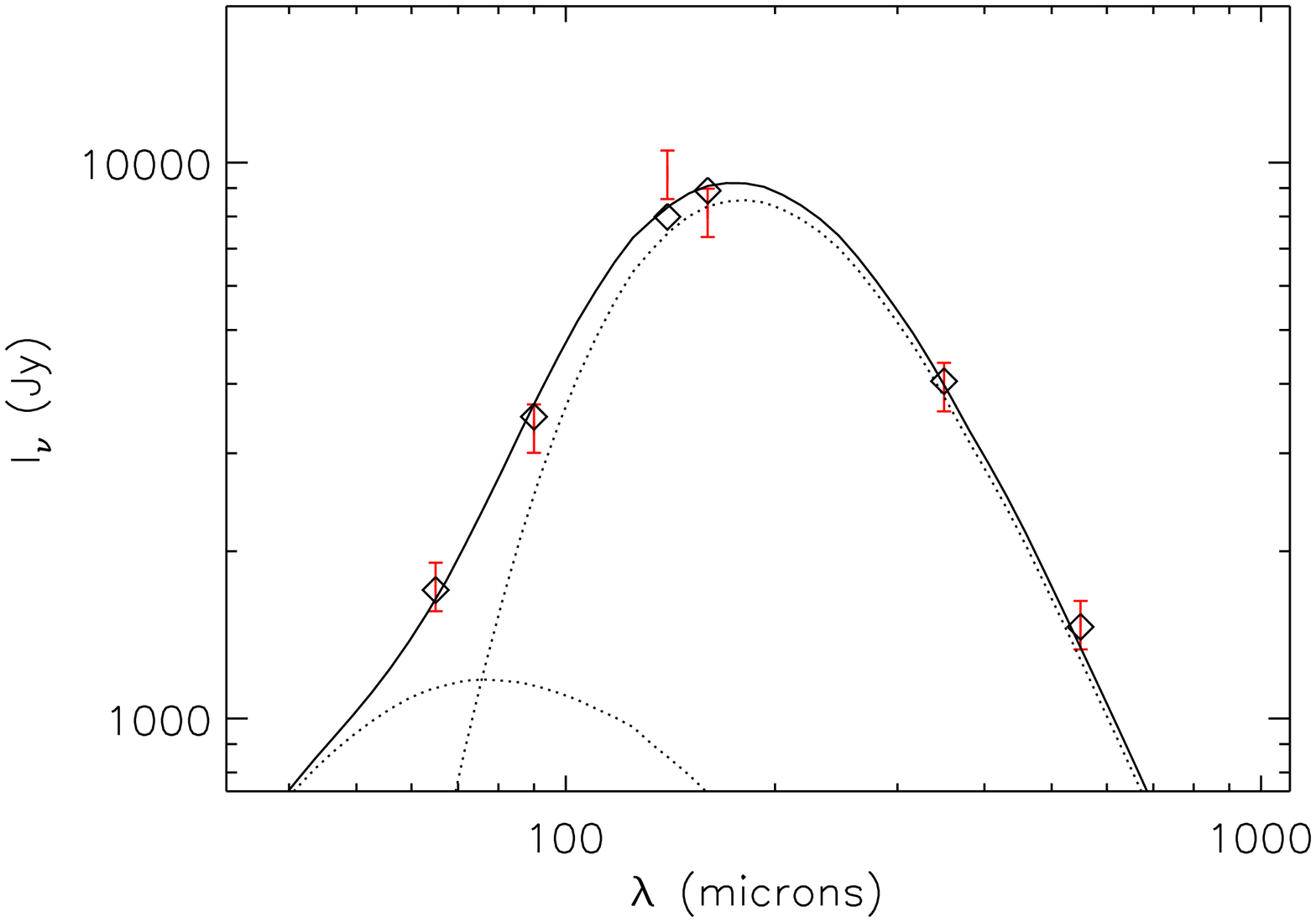}
\caption{SED of regions number 1 (\textit{upper left panel}), 2 (\textit{upper right panel}), 3 (\textit{lower left panel}), and 4  (\textit{lower right panel}). Diamonds indicate DUSTEM model prediction and fit. The red error bars denote photometric measurements at 65 $\mu$m, 90 $\mu$m, 140 $\mu$m, and 160 $\mu$m (AKARI), 350 $\mu$m, and 550 $\mu$m (Planck). The solid line shows the VSG+BG model, the dotted line at longer wavelength indicates the BGs, and the dotted line at shorter wavelength indicates the VSGs models.}
\label{fits}
\end{figure*}

Figure \ref{fits} shows  a section of the far-IR SED for  region 1 ({upper left panel}), region 2 ({upper right panel}), region 3 ({lower left panel}), and region 4 ({lower right panel}).
On this wavelength range one can find an equilibrium dust temperature set by
the thermal balance of the big dust grains and surrounding radiation field. Smaller dust grains
are stochastically heated, and therefore, it is meaningless to assign them a dust temperature
\citep[see e.g.][]{dra01}. For each region, the models provide the interstellar radiation field \citep[ISRF;][]{mat83}
needed for the dust to reach  thermal balance,
the total mass of dust (D\_M), and the temperatures for the three dust components that are assembled
at the wavelength range responsible for the infrared radiation, i.e. small carbon (smC), large carbon (laC), and silicates (Si). The  parameters derived from the SED fittings are listed in Table \ref{seds-parameters}.
In all four regions the ISRF is within a factor 2 of the mean standard value for the interstellar medium
and this indicates that the  involved radiation fields  are not strong. The dust temperature from the carbon
grains also reflect this trend with values close to 20 K, very representative of the more diffuse interstellar
medium \citep{bou96}. However, the temperature of the large silicates grains is indeed $\sim$5-6
degrees lower in three of the regions, and within the range of values expected for cold molecular 
gas \citep[see e.g.][]{fla09}  based on far-IR measurements. This mixture of temperatures is
unavoidable as we measure the flux densities along the line of sight.
The  temperatures derived for region 3 are
   systematically higher by 5-7  degrees. Although this
could be real, it may reflect some contamination of the
measurements by the presence of the Sh2-168 and 169
\hii\, regions that are found nearby along the line of sight.
 The small hump or bump in the dotted line at shorter wavelengths (see Fig. \ref{fits}, lower right panel) indicates that there is enough  contribution of VSGs to the SED  to correspond to around 10\% of the total mass of dust that is still dominated by BGs.

\begin{table}
\tiny
\caption{Parameters derived from the SEDs.} 
\label{seds-parameters}
\centering
 \begin{tabular}{l c c c c c}
\hline\hline
 Region& smC (K)& laC (K)& Si (K)& ISRF  &D\_M (M$_\odot$)\\
\hline
1 & 20.0     &  18.9    &   13.7   & 0.44  & 3.0$\times 10^3$\\
2 & 20.3     &  19.1    &   13.8   & 0.47  & 1.3$\times 10^3$\\
3 & 27.5     &  26.0    &   18.3   & 2.56  & 1.3$\times 10^2$\\
4 & 20.9     &  19.8    &   14.2   & 0.56  & 8.8$\times10^2$\\
\hline
\end{tabular}

\end{table}

\section{Origin of \g}\label{origin}

In this section, we analyse whether the origin of the supershell can be attributed to the action of stellar winds of massive stars and/or to the subsequent explosion as supernova. Along the narrow Galactic longitude range of about 10$^\circ$
from {\it l}$\sim$ 110$^\circ$ to  {\it l}$\sim$ 120$^\circ$, at least four 
 OB associations (Cas\,OB2, Cas\,OB5, Cep\,OB4, and Cas\,OB4) have
 been catalogued \citep{gar92}. All of these but Cep\,OB4 are located in the Perseus arm, and their heliocentric distances fall in the range from 1.8 to 2.8 kpc. 
 According to their mean Galactic longitude and latitude as derived by \cite{mel09}, only Cas\,OB5, $(l, b) = (116\fdg1, -0\fdg5)$, is located within the ellipse  delineating \g. These authors also
give a mean heliocentric distance of 2 kpc for  Cas\,OB5, which, bearing in mind the large errors involved in distances of OB associations, could be compatible with the distance derived for \g.
The fact that the mean Galactic coordinates  of Cas\,OB5  do not coincide with the centre of \g\  could be due to the presence of inhomogeneities of the interstellar medium and/or the possibility that some star members of Cas\,OB5 could have travelled from their hypothetical central position towards its present location owing to their peculiar spatial motions.

Although the long kinematic timescale of  about 11 Myr  estimated for \g\, 
suggests that more than one generation of massive stars
should be involved and/or that contributions from one or more SN explosion are to be expected, as a first step we carried out  energetic analysis to test whether the energy injected by the Cas\,OB5 stars lying inside \g\,  could have created the supershell.
Those stars are listed in Table \ref{cas}. Column 1 gives the stars
identification, Cols. 2 and 3 their Galactic coordinates, and Col. 4
their spectral types as given by \cite{gar92}.
We
estimated the main-sequence (MS) spectral type for each star
(Col. 5) based on the evolutionary track models published by \cite{sch92}, and adopting
the bolometric magnitudes and effective
temperatures given by \cite{gar92}. Column 6 gives an estimate of the star MS lifetime
(t(MS)) as derived from the stellar models of \cite{sch92}. The values given in this column are a rough estimate as a
consequence of the uncertainty in the mass adopted for each star.
Columns 7 and 8 give the mass loss rates ($\dot M$) and wind velocities
($V_w$) taken from \cite{lam93} and \cite{lei92}, respectively. Column 9
gives the total wind energy released by each star during its main-sequence
phase, $E_w = 0.5\,\dot M\, V^2_w$\,t(MS). Given that BD+61\,2554, BD+61\,2549,  BD+61\,2559, LSI+61\,112, BD+62\,2332, and HD\,108 are still in the MS, their values are upper limits. Assuming that these stars belong to Cas\,OB5, it is evident that  the star formation process in the association was not coeval and at least these six  stars may have been formed later.

On the other hand, taking  the total gaseous and molecular mass of \g\, derived in previous sections into account, the  kinetic energy stored in the shell can be calculated as $E_{\rm kin} = 0.5 (M_{\rm {t}}+M_{\rm {H_2}})\, v_{exp}^2 = (4.0 \pm 2.3) \times 10^{50}$ erg. 
From the last column of Table \ref{cas}, we infer that the total wind energy injected by the listed stars is $ E_w \sim\, 4.7 \times 10^{50}$ erg, which would only be sufficient to create \g\,  in the case in which the energy conversion efficiency from the wind energy into mechanical energy,  $\epsilon = E_{\rm kin}/E_w$,  were higher than  $\sim$ 85 \%.  This efficiency is too high even if the theoretical energy conserving model, which yields  $\epsilon = 0.2$,  were applicable \citep{wea77}. 

Therefore, an additional energy source seems to be needed to explain the supershell's origin.
If the origin of \g\, is indeed the action of many massive stars, for the sake of illustration we can estimate
 how many of them would have been 
needed to create it.
Taking $\epsilon = 0.2$ and adopting mean stellar wind parameters for O-type stars 
\citep{mok07} we find, for instance, that either four O6, or seven O7, or 83 B0 type stars are needed to create the structure.

However, since SNe are believed to play a predominant role in shaping the large-scale structure of the ISM \citep[e.g.][]{hen14} and the fact that the region under study involved several OB associations and large \hi\, shells, we cannot discard the possibility that  at least one supernova explosion had already taken place in the region.
To evaluate this possibility, we can make some rough estimations. 

According to \citet{cio88}  the maximum observable radius of an SNR before it merges
with the ISM  is given by
$$R_{\rm merge} = 51.3 \, E_0^{31/98} \, n^{-18/49}\, Z^{-5/98}\rm\, pc,$$
\noindent where $E_0$ is the explosion energy in units of $10^{51}$ erg,
$n$  the mean ambient particle density in $\rm cm^{-3}$, and
$Z$  the metallicity normalized to the solar value. 
The ambient density strongly depends on the Galactocentric distance ($R$) and the distance $z$ above the disk. To estimate the ambient density, we made use of \hi\, distribution model for the outer Milky Way  given by \citet{kal08}, which yields, for $7 \le R \le 35 $ kpc, the following radial distribution:
$$ n(R, z_0) \sim n_0\, exp^{-(R - R_{\odot})/R_n},$$
\noindent where $ n_0 = 0.9$ cm$^{-3}$,  $R_{\odot} = 8.5$ kpc, and $R_n = 3.15$ kpc is the radial scale length.
Assuming for \g\, a Galactocentric distance of $R = 10.3$ kpc, we obtain $n_0 \sim 0.5$ cm$^{-3}$.
Thus, for a solar metallicity and   $E_0 = 1,$ we obtain $R_{\rm merge} \sim 66$ pc, while for $E_0 = 2$, $R_{\rm merge}$ turns out to be about 82 pc. If we assume a lower metallicity value, $Z = 0.1$, we obtain  $R_{\rm merge} \sim 74 $  and $R_{\rm merge} \sim 93$ pc, for $E_0 = 1$ and $E_0 = 2$, respectively.
Thus, bearing in mind all the approximations used in these estimates, we believe that a scenario in which several OB stars  acted through their winds together with one or two SNe is plausible to explain the origin of \g.

\begin {table*}
\caption{Cas\,OB5 stars lying inside \g.} 
\label{cas}
\centering
\begin{tabular}{l c c c c c c c c c }
\hline\hline
Star & $ l$ & $b $ &Sp. Type & MS Sp. Type &  MS lifetime (Myr)& log ($\dot{M}$ (M$_{\odot} yr^{-1}$)) & $V_w$ (\kms) & $E_w$ ($10^{50}$ erg) \\
\hline
LSI+62\,28 & 115.9 & 0.7 & B1\,III & B1 & 26&-8.1 & 2100& 0.07\\
LSI+62\,32 & 115.9 & 0.4 & B2.5\,III & B1 &26 & -8.1&2100 & 0.07 \\
HD223767 & 116.0 & -0.2 & A4\, Iab & B1 &26 & -8.1&2100 & 0.07\\
BD+61\,2550 & 116.1 & 0.0 & O9.5\,II & O9  & 8&-7.2 & 2700 & 0.4 \\
BD+61\,2554 & 116.1 & 0.1 & B2\,V & B2 & 68& -8.8& 2200& $\leq$ 0.04\\
BD+61\,2549 & 116.2 &0.5 & B2\,V & B2 & 68& -8.8& 2200& $\leq$ 0.04\\
LSI+61\,92 & 116.2 & -0.4 & B2\,III  & B2&68 & -8.8& 2200&0.04\\
BD+61\,2559 & 116.3 &0.3 & O9\,V & O9 & 8& -7.2 &2700 & $\leq$ 0.4\\
BD+62\,2313 & 116.3 & 1.2 & B3\,IV & B1 & 26 &-8.1 &2100 & 0.07\\
HD224055 & 116.3 & -0.3  & B3\,Ia & O7 & 6.4&  -6.5 &2700 &1.4 \\
LSI+61\,98 & 116.4 & -0.1 & B3\,III & B1 & 26 &-8.1 & 2100& 0.07\\
LSI+61\,112 & 116.5 & -0.8 &B2\,V &B2 & 68&-8.8 &2200 &$\leq$ 0.04\\
BD+62\,2332 & 116.8 & 0.8 & B2\,V & B2 & 68& -8.8& 2200&$\leq$ 0.04\\
LSI+62\,2353 & 117.4 & 0.6 & B1\,III & B1 &26 & -8.1& 2100&0.07\\
HD225094 & 117.6 & 1.3 & B3\,Ia & O7 & 6.4 & -6.5 &2700 &1.4 \\
HD108* & 117.9 & 1.2 & O4-8\,f & O4-8 & 2.0 & -7.0&2000 &$\leq$ 0.1\\
BD+62\,1 &118.0 & 0.6 & B2\,IV & O9 & 8 & -7.2 &2700  &0.4\\
\hline
\end{tabular}
\tablebib{:* All the parameters of this star were obtained from \cite{mar12}}
\end{table*}

\section{Star formation}

To analyse whether star formation is still taking place in the region,
we attempt to identify young stellar object candidates (cYSOs) located in projection onto the molecular gas  possibly related to \g.
To this end, we made used of  the IRAS Point Source Catalogue \citep{bei88}, the MSX Infrared Point Source Catalogue \citep{ega03}, and the WISE  All-Sky Source Catalog \citep{wri10}.
Within a circular area of 3 degrees radius, centred at ($l, b$) = (117\fdg8, 1\fdg1), a total of 605 IRAS, 32 MSX, and 91191  WISE  sources  were found. From the MSX catalogue, we only considered those sources with acceptable flux quality  (q $\geq$ 2) in the four  bands, while  we selected only the sources with  a photometric uncertainty $\le$ 0.2 mag  and S/N $\ge$ 7 in all four bands for the WISE catalogue.

To identify the cYSOs among all the listed infrared sources, we applied colour criteria that are widely used to this end. 
In the case of the IRAS sources, 
we followed the \citet{jun92} colour criteria and select, among the 605 sources listed, 32 cYSOs. 
As for  the MSX sources, we applied the
 \citet{lum02} criteria, and found six massive young stellar objects (MYSO) candidates.
To classify the WISE sources, we adopted the  classification scheme described in \citet{koe12}, where the first step consists in removing various non-stellar  sources, such as PAH-emitting galaxies, broad-line active galactic nuclei (AGNs), resolved knots of shock emission, and resolved structured PAH-emission features
 from the listed sources. In this case, we removed a total of 46136  sources. Then, applying the colour criteria given by \citet{koe12} to the remaining sample of 45055 sources, we identified 148 Class I (sources where the IR emission arises mainly from a dense infalling envelope) and 385 Class II (sources where the IR emission is dominated by the presence of an optically thick circumstellar disk) cYSOs.

Bearing in mind that star formation takes place in molecular clouds, as an additional constraint we selected, from all the cYSOs found, only those lying in projection onto molecular gas observed in the velocity range from --53 to --40 \kms\, (see Fig. \ref{co}, \textit{\textup{upper panel}}).
Finally, after applying this constraint, the list of cYSOs contains 11 IRAS sources, 2 MYSOs, and 92 WISE sources, 32 Class I, and 60 Class II.

\begin{figure}
\includegraphics[scale=0.35]{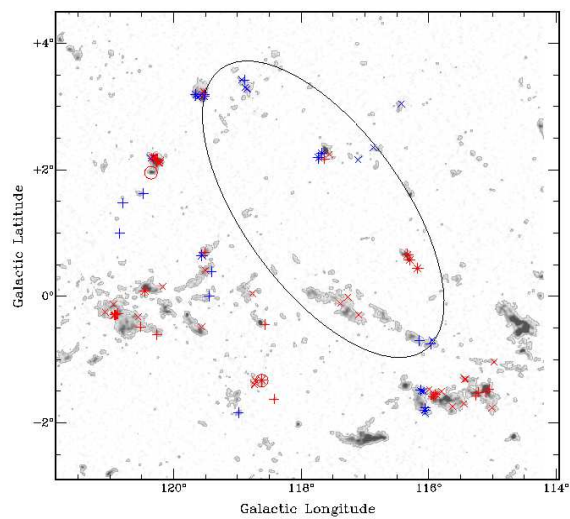}
\caption{Integrated CO emission distribution in the velocity range from  --53 to -- 40 \kms. Contour level is at 2.0 K. The location of the cYSOs  are indicated with different symbols according to the catalogue where they were found: MSX: circles; IRAS: asteriscs; WISE: plus signs correspond to Class I objects and crosses correspond to Class II. Red and blue  indicate objects in which direction there is just one CO component or more than one, respectively.   }
\label{ysos}
\end{figure}

Figure \ref{ysos} shows the integrated CO emission distribution in the velocity range considered with  the final list of cYSOs overlaid.  
Different symbols  indicates different classes of sources.
 It is important to mention that in the direction of 61 of the total cYSOs, CO emission is  only observed in the  velocity range from --53 to --40 \kms\, (they are indicated in red in Fig. \ref{ysos}), while in the direction of the rest, 31 sources the CO profile presents more than one component (these cYSOs are in blue  in Fig. \ref{ysos}).  
From Fig. \ref{ysos}, it is clear that the region is still undergoing star formation.

\section{Conclusions}

In this work we have studied in detail the structure labelled  \g\, in the supershell candidates catalogue  of \cite{sua14}. This feature is a well-defined \hi\, supershell centred at ($l, b$) = (117\fdg7, 1\fdg4), with an effective radius of $R_{ef} = 94 \pm 15$ pc, and a total gaseous and  molecular mass of $M_{\rm {t}} =  (4.9 \pm 2.2)\times 10^5 \,M_{\odot}$ and $M_{\rm{H_2}}=(7.4 \pm 3.2)\times 10^4$ M$_\odot$, respectively.

Diffuse radio continuum emission is detected around the supershell. A study of the temperature spectral index distribution reveals that thermal emission is only found  in a small region, located towards the eastern boundary of \g, while  non-thermal emission dominates the remaining parts.
We have also compared the morphology of the HI supershell with that observed
in the mid- and far-IR and conclude that they are quite similar, in particular, in the range of \hi\, velocities
from $-41.9$ to $-47.6$ \kms,  suggesting that dust components coexist with the \hi\, gas. Based on this good morphological coincidence, we  used far-IR measurements to determine the dust temperature of four regions located
around the supershell. For three of these regions, we  found   a mean dust temperature of  19.3 K, as traced 
   by the Big Grains, not far from that  expected in molecular gas. The fourth region is a bit warmer and closer in temperature to what is expected for the diffuse interstellar
medium.

 We conclude, based on an energetic analysis, that the origin of \g\, could be attributed to the mechanical energy injected by the winds of the star members of Cas\,OB5 observed  projected towards the inner part of the shell in combination with another energy source, such as the stellar winds of other massive stars and/or one or two SN explosion. This result is in agreement with the predominance of non-thermal emission found around \g.

We have also analysed the possible association of several \hii\, regions and three SNRs that are seen  superimposed
   on the line-of-sight radio continuum maps and conclude that, based on their known distances or radial velocities, none of the SNRs and some of the \hii\, regions are physically related to the  \hi\, supershell.
    On the other hand, four of the \hii\, regions may be at the same distance of \g. The fact that these four \hii\, regions are projected onto the supershell borders leads us to the conclusion that they may have been created as a consequence of the action of a strong shock produced by the expansion of \g\, into the surrounding molecular gas.

In the region under study there are several small CO clouds observed around the supershell that could be either the debris of the original CO cloud out of which  Cas\,OB5  was  created or they were created in the dense walls of \g.

All the evidence found in this mutliwavelength study support the idea that \g\,  is actually a physical structure with dense and molecular gas. On the other hand, 
the presence of several cYSOs projected onto the molecular clouds and four \hii\, regions  detected onto the supershell allow us to conclude that the area is still an active star-forming region. 

\begin{acknowledgements} 

We thank the referee for her/his careful reading and comments on the manuscript
that has helped us to improve its content and presentation. The CGPS is a Canadian Project with international partners and is supported by grants from NSERC. Data from the CGPS are publicly available through the facilities of the Canadian Astronomy Data Centre (http://cadc.hia.nrc.ca) operated by the Herzberg Institute of Astrophysics, NRC. This project was partially financed by the Consejo Nacional de Investigaciones Cient\'{i}ficas y T\'ecnicas (CONICET) of Argentina under project PIP 01299, Agencia PICT 00902, and UNLP G091. L.A. Suad is a post-doctoral fellow of CONICET, Argentina. S. Cichowolski and E.M. Arnal are members of the \textit{Carrera del Investigador Cient\'{i}fico} of CONICET, Argentina. J.C. Testori is member of the \textit{Carrera del Personal de Apoyo}, CONICET, Argentina.

 \end{acknowledgements}

\bibliographystyle{aa} 
\bibliography{bibliografia}
  
 \IfFileExists{\jobname.bbl}{}
{\typeout{}
\typeout{****************************************************}
\typeout{****************************************************}
\typeout{** Please run "bibtex \jobname" to optain}
\typeout{** the bibliography and then re-run LaTeX}
\typeout{** twice to fix the references!}
\typeout{****************************************************}
\typeout{****************************************************}
\typeout{}
}

\end{document}